

\input harvmac.tex
\noblackbox

\lref\taylor{T.R. Taylor, ``Dilaton, Gaugino Condensation and
Supersymmetry Breaking,'' {\it Phys. Lett.} {\bf B252} (1990) 59.}
\lref\nilles{J.P. Derendinger, L.E. Ibanez, and H.P. Nilles, ``On the Low
Energy D=4, N=1 Supergravity Theory Extracted from the D=10, N=1
Superstring,'' {\it Phys. Lett.} {\bf 155B} (1985) 65.}
\lref\sd{N. Seiberg, ``Electric-Magnetic Duality in Supersymmetric Nonabelian
Gauge Theories,'' {\it Nucl. Phys.} {\bf B435} (1995) 129.}
\lref\othmath{R. Kobayashi and A.N. Todorov, ``Polarized Period Map
for Generalized K3 Surfaces and the Moduli of Einstein Metrics,''
{\it Tohoku Math. J.} {\bf 39} (1987) 341 \semi
M.T. Anderson, ``The $L^{2}$ Structure of Moduli Spaces of Einstein Metrics
on 4-Manifolds,'' {\it Geom. and Funct. Analysis} {\bf 2} (1992) 29.}
\lref\asp{P. Aspinwall, ``Enhanced Gauge Symmetries and K3 Surfaces,''
{\it Phys. Lett.} {\bf B357} (1995) 329.}
\lref\Schimmrigk{M. Lynker and R. Schimmrigk, ``Conifold Transitions
and Mirror Symmetries'', hep-th/9511058.}
\lref\Aspinwall{P. Aspinwall, ``An N=2 Dual Pair and a Phase Transition,''
hep-th/9510142.}
\lref\BSV{M. Bershadsky, V. Sadov, and C. Vafa, ``D Strings
on D Manifolds'', hep-th/9510225.}
\lref\HT{C. Hull and P. Townsend, ``Unity of Superstring Dualities,'' {\it
Nucl. Phys.} {\bf B438} (1995) 109 \semi
E. Witten, ``String Theory Dynamics in Various Dimensions,'' {\it Nucl. Phys.}
{\bf B443} (1995) 85.}
\lref\Ibanez{A. Font, L. Ibanez, D. Lust, and
F. Quevedo, ``Strong-Weak Coupling
Duality and Nonperturbative Effects in String Theory'',
{\it Phys. Lett.} {\bf B249} (1990) 35\semi
J. Horne and G. Moore, ``Chaotic coupling constants'', {\it Nucl. Phys.}
{\bf B432} (1994) 109.}
\lref\Ald{G. Aldazabal, A. Font, L. Ibanez, and F. Quevedo,  ``Chains of N=2,
D=4 Heterotic Type II Duals'', hep-th/9510093.}
\lref\hoslian{S. Hosono and B. Lian, unpublished.}
\lref\hubsh{P. Candelas, A. Dale, A. Lutken, and R. Schimmrigk,
``Complete Intersection Calabi-Yau Manifolds'',
{\it Nucl. Phys.} {\bf B}298 (1988) 493.;
P. Green and T. Hubsch, ``Connecting Moduli Spaces
of Calabi-Yau Threefolds'', {\it Commun. Math. Phys.} 119 (1988) 431;
``Phase Transitions among (Many of) Calabi-Yau Compactifications'',
{\it Phys. Rev. Lett.} {\bf 61} (1988) 1163\semi
P. Candelas, P. Green, and T. Hubsch, ``Finite Distances Between Distinct
Calabi-Yau Vacua,'' {\it Phys. Rev. Lett.} {\bf 62} (1989) 1956;
``Rolling Among Calabi-Yau Vacua,'' {\it Nucl. Phys.} {\bf B330} (1990) 49.}
\lref\GMS{B. Greene, D. Morrison, and A. Strominger, ``Black
Hole Condensation and the Unification of String Vacua'',
{\it Nucl. Phys.} {\bf B}451 (1995) 109.}
\lref\strom{A. Strominger, ``Massless Black Holes and
Conifolds in String Theory'', {\it Nucl. Phys.} {\bf B}451 (1995) 96.}
\lref\KV{S. Kachru and C. Vafa, ``Exact Results for N=2 Compactifications
of Heterotic Strings'', {\it Nucl. Phys.} {\bf B450} (1995) 69.}
\lref\FHSV{S. Ferrara, J. Harvey, A. Strominger, and
C. Vafa, ``Second Quantized Mirror Symmetry'', {\it Phys. Lett.}
{\bf B361} (1995) 59.}
\lref\nonkahlerres{See e.g. R. Friedman, ``On Threefolds with Trivial
Canonical Bundle,'' {\it Complex Geometry and Lie Theory}, J. Carlson,
H. Clemens, and D. Morrison, eds., {\it Proceedings of Symposia in
Pure Mathematics} {\bf 53} AMS, (1991) 103-135.}
\lref\NIIrefs{
V. Kaplunovsky, J. Louis, and S. Theisen, ``Aspects of Duality in
N=2 String Vacua,'' {\it Phys. Lett.} {\bf B357} (1995) 71\semi
I. Antoniadis, E. Gava, K.S. Narain, and T.R. Taylor, ``N=2 Type II
Heterotic Duality and Higher Derivative F Terms,'' hep-th/9507115\semi
S. Kachru, A. Klemm, W. Lerche, P. Mayr, and C. Vafa, ``Nonperturbative
Results on the Point Particle Limit of N=2 Heterotic String
Compactifications,'' hep-th/9508155\semi
I. Antoniadis and H. Partouche, ``Exact Monodromy Group
of N=2 Heterotic Superstring,'' hep-th/9509009.
}
\lref\monod{M. Billo et al., ``A Search for Non-perturbative Dualities of
Local N=2 Yang-Mills Theories from Calabi-Yau Threefolds,'' hep-th/9506075\semi
G. Cardoso, D. Lust, and T. Mohaupt, ``Nonperturbative Monodromies in N=2
Heterotic String Vacua,'' hep-th/9507113.}
\lref\joe{J. Polchinski, ``Dirichlet Branes and Ramond-Ramond Charges'',
hep-th/9511026.}
\lref\Lowe{J. Harvey, D. Lowe, and A. Strominger, ``N=1 String Duality'',
hep-th/9507168.}
\lref\VW{C. Vafa and E. Witten, ``Dual String Pairs with N=1
and N=2 Supersymmetry in Four-Dimensions'', hep-th/9507050.}
\lref\SW{N. Seiberg and E. Witten, ``Electric-Magnetic Duality,
Monopole Condensation, and Confinement in N=2 Supersymmetric
Yang-Mills Theory'', {\it Nucl. Phys.} {\bf B}426 (1994) 19,
Erratum, {\it ibid.} {\bf B}430 (1994) 485.}
\lref\edthree{E. Witten, ``Strong Coupling and the Cosmological
Constant'', {\it Mod. Phys. Lett.} {\bf A10} (1995) 2153.}
\lref\ussol{S. Kachru and E. Silverstein, ``Nonsupersymmetric String
Solitons'', hep-th/9508096, to appear in {\it Nucl. Phys.} {\bf B}.}
\lref\Pouliot{P. Pouliot, ``Duality in SUSY SU(N) with
an Antisymmetric Tensor Field'', hep-th/9510148\semi
P. Pouliot and M. Strassler, to appear
\semi K. Intriligator and S. Thomas, to appear. }
\lref\sun{A. Klemm, W. Lerche, S. Yankielowicz, and S. Theisen,
``Simple Singularities and N=2 Supersymmetric Yang-Mills Theory,''
{\it Phys. Lett.} {\bf B344} (1995) 169 \semi P. Argyres and A. Faraggi,
``The Vacuum Structure and Spectrum of N=2 Supersymmetric SU(N) Gauge
Theory,'' {\it Phys. Rev. Lett.} {\bf 74} (1995) 3931.}
\lref\son{U. Danielsson and B. Sundborg, ``The Moduli Space and
Monodromies of N=2 Supersymmetric SO(2r+1) Yang-Mills Theory,''
hep-th/9411048\semi A. Brandhuber and K. Landsteiner,
``On the Monodromies of N=2 Supersymmetric Yang-Mills Theory with
Gauge Group SO(2N),'' hep-th/9507008.}
\lref\fib{A. Klemm, W. Lerche, and P. Mayr, ``K3 Fibrations and
Heterotic Type II String Duality,'' {\it Phys. Lett.} {\bf B357}
(1995) 313\semi
P. Aspinwall and J. Louis, ``On the Ubiquity of K3 Fibrations in
String Duality,'' hep-th/9510234.}
\lref\wilsrefs{L. Ibanez, H. P. Nilles, and F. Quevedo,
{\it Phys. Lett.} {\bf B187} (1987) 25.}
\lref\SIrev{K. Intriligator and N. Seiberg, ``Lectures on Supersymmetric
Gauge Theory and Electric-Magnetic Duality'', hep-th/9509066.}
\lref\gaugecond{See e.g. D. Amati, K. Konishi, Y. Meurice, G.C. Rossi, and
G. Veneziano, ``Nonperturbative Aspects in Supersymmetric Gauge Theories,''
{\it Phys. Rep.} {\bf 162} (1988) 169.}

\lref\drsw{M. Dine, R. Rohm, N. Seiberg, and E. Witten, ``Gaugino
Condensation in Superstring Models'',
{\it Phys. Lett.} {\bf B156}
(1985) 55.}
\lref\krasnikov{N. Krasnikov, ``On Supersymmetry Breaking in
Superstring Theories,''
{\it Phys. Lett.} {\bf B193}
(1987) 37. }
\lref\dixrev{L. Dixon, ``Supersymmetry Breaking in String Theory'',
talk presented at APS DPF meeting in Houston, Jan 3-6 (1990).}
\lref\HS{J. Harvey and A. Strominger, ``The Heterotic String
is a Soliton'', {\it Nucl. Phys.} {\bf B}449 (1995) 535 \semi
A. Sen, ``String-String Duality Conjecture in Six-Dimensions
and Charged Solitonic Strings'', {\it Nucl. Phys.} {\bf B450}
(1995) 103.}
\lref\cdgp{P. Candelas, X. de la Ossa, P. Green, and L. Parkes,
``A Pair of Calabi-Yau Manifolds as an Exactly Solvable
Superconformal Field Theory'', {\it Nucl. Phys.} {\bf B359}
(1991) 21.}
\lref\PS{J. Polchinski and A. Strominger, ``New Vacua for Type II String
Theory'', hep-th/9510227.}
\lref\Nik{V. Nikulin, ``Finite Automorphism Groups of Kahler
K3 Surfaces'', {\it Trans. Moscow Math. Soc.} {\bf 38} (1980) 71.}
\lref\dave{D. R. Morrison, ``Some Remarks on the Moduli of K3 surfaces'',
{\it Classification of Algebraic and Analytic Manifolds}, K. Ueno, ed.,
{\it Progress in Math.} {\bf 39} Birkhhauser, Boston, Basel, Stuttgart,
(1983) 303-332.}
\lref\edcomments{E. Witten, ``Some Comments on String Dynamics'',
hep-th/9507121.}
\lref\phases{E. Witten, ``Phases of N=2 Theories in Two Dimensions'',
{\it Nucl. Phys.} {\bf B}403 (1993) 159.}
\lref\silvwitt{E. Silverstein and E. Witten, ``Criteria for Conformal
Invariance of (0,2) Models'', {\it Nucl. Phys.} {\bf B}444 (1995) 161.}
\lref\carlosetal{B. de Carlos, J. Casas, C. Munoz, ``Supersymmetry
Breaking and Determination of the Unification Gauge Coupling Constant
in String Theories'', {\it Nucl. Phys.} {\bf B399} (1993) 623.}
\lref\kaplou{V. Kaplunovsky and J. Louis, ``Field Dependent Gauge
Couplings in Locally Supersymmetric Effective Quantum Field
Theories'', {\it Nucl. Phys.} {\bf B}422 (1994) 57.}

\Title{\vbox{\hbox{HUTP-95/A046}\hbox{PUPT-1579}\hbox{\tt hep-th/9511228}}}
{\vbox{\centerline{N=1 Dual String Pairs and Gaugino Condensation}}
        }
\centerline{Shamit Kachru\footnote{$^*$}
{kachru@string.harvard.edu}}
\smallskip\centerline{\it Lyman Laboratory of Physics}
\centerline{\it Harvard University}\centerline{\it Cambridge, MA 02138}
\bigskip
\medskip
\centerline{Eva Silverstein\footnote{$^\dagger$}
{silver@puhep1.princeton.edu}}
\smallskip\centerline{\it Joseph Henry Laboratories}
\centerline{\it Jadwin Hall}
\centerline{\it Princeton University}\centerline{\it Princeton, NJ 08544}
\vskip .2in


We study a class of four-dimensional N=1 heterotic string
theories which have nontrivial quantum dynamics arising from
asymptotically free gauge groups.  These models are obtained
by orbifolding 4d N=2 heterotic/type II
dual pairs by symmetries which leave unbroken products of
nonabelian gauge groups
(without charged matter) in a ``hidden sector''  on the heterotic side.
Such models are expected to break supersymmetry
through gaugino condensation in the hidden sector.  We find
a dual description of the effects of gaugino condensation
on the type II side, where
the corresponding superpotential arises at tree level.
We speculate that the conformal field theory underlying the type II
description may be related to
a class of geometrical nonsupersymmetric string
compactifications.

\Date{November 1995} 

\newsec{Introduction}

There has been much progress in recent months in unifying
various string theories in different dimensions using the
appropriate versions of strong/weak coupling duality.
In compactifications with enough supersymmetry,
the
low-energy physics is more or less determined by the constraints of
supersymmetry.
In compactifications with the equivalent of $N\le 2$ supersymmetry
in four dimensions, there emerges the possibility of nontrivial
dynamics in the infrared.
While the most famous example of ``string-string'' duality between
heterotic and type II strings involves theories with the equivalent of
N=4 supersymmetry in four dimensions \HT\HS, more recently
dual heterotic/type II pairs with $d=4, N=2$ supersymmetry have been
discovered
\KV\FHSV\ and studied in some detail
\refs{\NIIrefs,\monod,\fib,\VW,\Ald,\Aspinwall}.  One
finds that the dual description renders
instanton effects computable using classical string theory.

The theories of interest for describing low-energy particle
physics have $N\le 1$ supersymmetry in four dimensions.
In particular, there has been much interest in
$4d$ $N=1$ heterotic models for string phenomenology.
One mechanism for supersymmetry breaking that has been much
studied in this context is gaugino condensation in a
hidden sector \refs{\nilles,\drsw,\krasnikov,\dixrev,\taylor,
\kaplou,\carlosetal}.

Starting from the $N=2$ dual pairs, one can form $N=1$ dual
pairs by freely acting orbifolds \VW\Lowe.  Unfortunately, the N=1
pairs constructed to date have had trivial dynamics in the infrared.
There are several different classes of infrared dynamics,
parametrized by the spectrum of gauge fields and charged matter present
in the ultraviolet,
that have been studied fruitfully in supersymmetric field theory (see
\SIrev\ for a review).
Understanding the
quantum behavior of N=1 string vacua in four dimensions
will involve learning
how string theory recovers and generalizes these phenomena.

In this paper we will study the role of duality in
elucidating the effects of gaugino condensation
for a class of examples.
Specifically, it is possible to construct
freely acting orbifolds producing N=1 dual pairs
which on the heterotic side have pure factors in the gauge
group.  Of particular interest are models with more
than one simple factor in the pure gauge group, as
some of these models are expected to lead to stable
vacua with broken supersymmetry \krasnikov\dixrev\taylor.
We describe how to construct such models in \S2.

On the type II side, this appears mysterious at first sight.
The singularities of the vector
multiplet moduli space of the $N=2$ theory
do not lead to nonabelian gauge symmetry enhancement (except in the case
of nonasymptotically free theories \BSV).  Therefore,
in the type II $N=1$ orbifold,
there is no obvious origin of quantum supersymmetry breaking
effects.  This suggests that the supersymmetry breaking, if
present, should be
evident classically.
Said differently, despite the lack of nonrenormalization
theorems to guarantee the utility of strong/weak coupling
duality in $N=1$ string theory, the absence of nonabelian dynamics on the
type II side essentially requires its description of
the physics to be perturbative, leading to a useful duality.
We will propose a tree level mechanism on the type
II side which reproduces the expected potential generated by
gaugino condensation in \S3.

The mechanism can be summarized as follows.
In the global limit it reduces to that which
Seiberg and Witten used for understanding the mass gap of pure $N=1$
gauge theory by perturbing pure $N=2$ gauge theory with a mass term
for the adjoint scalar $\phi$ \SW.
In their scenario turning on a mass perturbation leads to
a vacuum expectation values for the light monopole fields present at the
special singularities of the N=2 moduli space.
In string theory it is not possible to turn on masses by
hand.  We find nevertheless that the orbifold\foot{Actually on the type IIA
side we will be discussing an orientifold.} spectrum contains
a massive field, which has the same global couplings as the $\phi$
field studied in \SW, and which becomes light as the heterotic coupling
becomes weak.  In \S3 we explain how we can infer the presence and couplings
of such fields.
We then
study the bosonic potential in the resulting low-energy supergravity
theory and compare to the expectations of gaugino condensation on
the heterotic side.  In \S4 we recap and discuss directions
for further exploration of these models.

The purpose of this paper is to demonstrate how the
type II side manages to encode perturbatively the physics of the
heterotic side.  We wish to note here that the utility
of the proposed duality comes from the opposite approach.
The perturbative physics of the type II side should contain
a wealth of information
about the quantum behavior of the heterotic side, and
is a promising starting point for a systematic study of
the details of such models.
There has been other recent work on supersymmetric/nonsupersymmetric
duality in
field theory \Pouliot\ and
string theory \edthree\ussol.  Earlier attempts to use strong/weak
coupling duality
to shed light on gaugino condensates (by assuming an $SL(2,Z)$ S-duality
acting on the dilaton-axion multiplet) can be found in \Ibanez.

\newsec{Construction of Dynamical Duals}

Our starting point is the adiabatic prescription introduced
by Vafa and Witten for producing
dual string pairs from known examples of string-string duality.
The $N=2$ compactifications have the structure of
$K3$ fibrations (${\bf CP}^1$,$K3$)
on the type II side \fib\ and $T^4$ fibrations (${\bf CP}^1$,$T^4$)
on the heterotic side \VW.  Act with a freely
acting symmetry (in our case $(z_1,z_2)\rightarrow (\bar z_2,-\bar z_1)$) on
the base and approach the large radius limit of the ${\bf CP}^1$. Then
the local observer sees the physics of a compactification
on $K3$ or $T^4$ (on the type II and heterotic sides, respectively)
and maps reliably to the dual theory using the well-established
$6d$ string-string duality \HT\HS.

We will now discuss the class of
orbifolds of interest to us, first in the heterotic
and then in the type II description.

\subsec{The Heterotic side}

In order to apply the adiabatic argument of \VW, we look for
freely acting symmetries with which to orbifold the $N=2$ examples
of \KV.  On the heterotic $K3\times T^2$,
we can use the $Z_2$ given by
the Enriques involution on the
$K3\sim ({\bf CP}^1,T^2)$ and the reflection $-1$ on the $T^2$ \FHSV\VW.

Of course, we must also choose an embedding of the $Z_{2}$ in the
gauge degrees of freedom.
This will determine the
surviving gauge group and charged matter content.
We can obtain $N=1$ models with
pure factors in the gauge group as follows, starting from models 6-8 in
\S4.5 of \KV.  For definiteness, consider model 7, obtained
on the heterotic side by embedding an $SU(2)$
bundle with $c_2=20$ in one of the two $E_8$ factors
(which we will call $E_8^{{\rm obs}}$), breaking
it to $E_7$, and embedding a rank two bundle with
$c_2$=4 into an enhanced $SU(2)$ arising from the $T^2$ fixed
at $\tau=\rho$.  The generic
spectrum of 11 vectors and 377 hypermultiplets is obtained as
the Higgs phase of the $E_7$ and the Coulomb phase of the
second $E_8$ (which we will call $E_8^H$).
This model has a known type II Calabi-Yau dual described in \KV.

If we consider the heterotic side $\it before$ Higgsing the $E_7$, then
we see that
any further orbifolding to get an $N=1$ model
with modular-invariant embedding of the
orbifold group only into $E_8^{{\rm obs}}$ will
produce a model with a hidden sector.
Here we should emphasize that although the $N=2$ dual pair of \KV\
involves the heterotic model in which the $E_7^{obs}$
is completely Higgsed,
one should be able to follow both sides of the duality
through appropriate ``extremal transitions'' to the model with the $E_7$
unbroken
(examples of such transitions have been discussed in detail in
\GMS\Ald\Aspinwall\Schimmrigk).
In general there is no guarantee that the appropriate type
II dual will still be a Calabi-Yau compactification, but for
our example there $\it is$ in fact a good candidate for the ``unHiggsed''
dual, as we will explain presently.

The model we are discussing has a generic
spectrum -- before Higgsing the observable $E_7$ -- of
62 hypermultiplets and 18 vector multiplets.  This would correspond
to a type IIA string compactification on a Calabi-Yau with Hodge
numbers $h_{11}=17, h_{21}=61$.  There is indeed
a known $K3$ fibration
with these Hodge numbers,
given by the hypersurface of degree 68
in ${\bf WP}^{4}_{3,3,8,20,34}$ \hoslian.
The type IIA compactification on this manifold is a good candidate for the
dual of our ``unHiggsed'' heterotic theory.

Now that we have a candidate dual
for the heterotic theory with $E_8\times E_7\times
U(1)^3 $ gauge group (and with 8 ${\bf {56}}$s of $E_7$) we need to decide
which sorts of free group actions we want to orbifold by to obtain
interesting N=1 models.
Of more interest for phenomenology than the models with $E_8^{H}$ unbroken
are models with several
pure nonabelian factors in the gauge group \krasnikov.
We can obtain such models by first turning on discrete
Wilson lines (consistent with the above $Z_2$ action on the $T^2$)
to break $E_8^H$ to a subgroup and then orbifolding by the
$Z_2$.\foot{In turning on Wilson lines in model 7 of \KV, we
simultaneously move away from $\tau=\rho$ in such a way
as to preserve, in the presence of the Wilson lines,
the enhanced $SU(2)$ in which we have embedded a $c_2=4$ $SU(2)$
bundle}
Equivalently, this can be described as embedding
translations generating the $T^2$, which are part of the space group
of the orbifold, into $E_8^H$.  This procedure is constrained
by level-matching and by the relations of the space group
\wilsrefs.

There are several consistent choices which produce product hidden
sector groups.  For example,
one choice that leads to a hidden
sector gauge group $G^H=SO(8)^2$ is obtained
as follows.  Take Wilson lines $A_1=L_1/2$ and $A_2=L_2/2$ around
the two cycles of the $T^2$, where $L_1=(0,0,0,0,1,1,1,1)$ and
$L_2=(-2,0,0,0,0,0,0,0)$ are vectors in the $E_8$ root lattice.
Since $A_1^2$, $A_2^2$, and $A_1\cdot A_2$ are integers, this
satisfies level-matching.  With Wilson lines turned on, the
momentum lattice becomes
\eqn\momL{p_L=(P^I-A^I_in^i,G^{ij}({m_j\over 2}+{A^I_j\over 2}P^I
-B^{ij}n_j-{{A^I_iA^I_j}\over 4}n^j)+n^i)}
\eqn\momR{p_R=G^{ij}({m_j\over 2}+{A^I_j\over 2}P^I
-B^{ij}n_j-{{A^I_iA^I_j}\over 4}n^j)-n^i}
This shows that the surviving gauge bosons have root vectors satisfying
$A_i^IP^I\in{\bf Z}$.  The $E_8$ root lattice consists of
vectors of the form $\pm e_i\pm e_j$ and
${1\over 2}(\pm e_1\pm e_2\dots\pm e_8)$.  All of the first set
and none of the second set satisfy $A_2\cdot P\in {\bf Z}$.
There are 48 vectors in the first set which satisfy
$A_1\cdot P\in {\bf Z}$.  These split up into two copies
of the root lattice of $SO(8)$.  Combined with the 8 Cartan generators,
this gives the dimension 56, rank 8 group $G^H=SO(8)\times SO(8)$.
Other hidden sector product groups can be obtained similarly:  The
discussion of the type II side below applies to
all models of this type.

\subsec{The Type IIA side}

The heterotic orbifold was constructed by a $Z_2$ which acted
freely on the base (${\bf CP}^1$) of the elliptic fibration
as follows:
\eqn\acbase{(z_1,z_2)\rightarrow (\bar z_2,-\bar z_1).}
As we mentioned in \S2.1,
the $N=2$ heterotic model has a proposed dual, type IIA string
theory compactified
on the Calabi-Yau hypersurface in ${\bf WP}^4_{3,3,8,20,34}$
\Schimmrigk\hoslian.
This manifold is a $K3$ fibration, with the $K3$ fiber being
given by a degree 34 hypersurface in
${\bf WP}^3_{3,4,10,17}$.

By the adiabatic argument, we expect a type II dual of
the $N=1$ heterotic orbifold model which is obtained by translating the action
of the $Z_2$ on the heterotic coordinates to an action
on the harmonic forms of the $K3$ fiber \HS\VW.
On the heterotic side the orbifold left us with a pure
gauge factor, projecting out the charged fields which
in the N=2 theory parameterized the Coulomb phase
of $E_8^H$.  The singularities corresponding to
enhanced gauge symmetry on the heterotic side map
to the conifold singularities of the Calabi-Yau moduli
space on the type II side \KV\FHSV.
The corresponding orbifold
on the type IIA side will project out the scalars, $a_{b,D}^i$
$i=1,\dots,{\rm rank}(G_b^H)$.
These are the moduli which could move us away from the conifold locus dual
to the enhanced hidden
gauge symmetry locus of the heterotic string side.  On the other
hand, the abelian vectors in $N=2$
vector multiplets will survive the orbifold
projection.

Before embarking on an analysis of the physics of the dual descriptions,
we should make sure that the heterotic orbifold indeed maps
to a bona fide orbifold on the type II side.
That is, does the action $G$ on the harmonic forms of
$K3$ implied by the action on the Narain lattice on
the heterotic side \HS\ determine a symmetry of
the $K3$ itself?  More precisely, we must ensure that there is a symmetry
of the worldsheet action on the type II side by which we can orbifold.
The action on the base \acbase\ indicates that the symmetry
is antiholomorphic and orientation reversing which means
that we must also exchange left and right movers on the worldsheet,
giving us an orientifold as in the examples of \VW.
The novelty in our construction is that we must also find a symmetry
which freezes the $K3$ fibers at a singular $K3$, dual to the enhanced
gauge symmetry present in the heterotic orbifold \HT\asp.

If the heterotic theory is developing an enhanced gauge symmetry group
$G$, then
the dual description involves a $K3$ in which curves $C_i$ ($i=1,\cdots,
{\rm rank(G)}$) are shrinking to zero size (and the associated worldsheet
$\theta$
angles are also vanishing \asp).   In general, given a smooth
rational curve $C$, it satisfies $C \cdot C = -2$
(where $\cdot$ denotes the intersection product between
homology classes), so one can
consider an automorphism of $H^{2}(K3)$ given by the ``reflection''
\eqn\cohact{X \rightarrow X + (X \cdot C) C~}
which takes $C\rightarrow -C$.
This automorphism is $\it not$ associated with a symmetry of
the $K3$ --
to explain what it $\it is$ associated with, we need to recall and extend
the notion of a birational transformation.
\foot{We
thank D. Morrison for very helpful discussions about this and the following.}

Our extended notion of birational transformation will be as follows:
A birational transformation between $X$ and $Y$, both of dimension $d$,
is an algebraic cycle $Z \subset X\times Y$, also of dimension $d$,
such that for appropriate dense open subsets $U\subset X$ and
$V\subset Y$ the intersection of $Z$ with $U\times V$ is the graph of
an isomorphism.  Any such cycle induces a map $H^{k}(X) \rightarrow H^{k}(Y)$
in a natural way.  The important point for us, however, is that $Z$ is
allowed to contain more than one component: All but one of the components
will map to proper subvarieties in both $X$ and $Y$, and so will be disjoint
from $U\times V$.

In intuitive terms, this extension of the notion of birational transformation
has the following effect.  As far as complex structures are concerned, any
birational spaces $X$, $Y$ differ only at complex co-dimension two.  That is,
one can extend the isomorphism given by $U\times V$ to hold up to
codimension two, in a manner consistent with the complex structures of
$X$ and $Y$.  However, in string theory, we are also interested in keeping
track of the Kahler classes, and these can obstruct the extension of
the isomorphism even to codimension one subspaces of $X$ and $Y$.

So in fact, \cohact\ is the action on the cohomology induced by a
birational transformation in the sense discussed above.
More generally, one can define such a ``reflection'' associated with
any set of rational curves $C_i$ (see the discussion in \S3 of \dave\ and
also in \othmath).  These reflections are also associated with
birational transformations (in this extended sense) between distinct $K3$s.
The ``extra'' components of $Z$ will be $CP^{1}\times CP^{1}$s given by
the curves $C_i$ in the two $K3$s.
It turns out that the string compactifications on the two $K3$s are
isomorphic, however, so these ``reflections'' generate $Z_{2}$s acting on the
Teichmuller space for the moduli space of string theories on $K3$.
Heuristically speaking, one should envision a cone divided into
two ``mirror image'' cones $A$ and $B$ by a dividing wall in the center.
$A$ is the Kahler cone for one $K3$, and the $K3$ obtained by doing the
reflection \cohact\ (and the associated birational transformation)
instead has Kahler cone $B$.  The wall dividing the
two cones $A,B$ is the wall where the $C_i$ shrink to zero area.

It is well known that at the $\it fixed~ point$ of a symmetry group $g$
which acts on the Teichmuller space, one
obtains a conformal theory with an enhanced
$g$ symmetry.  So at the fixed point of the $Z_{2}$ generated by such
a reflection, one will find a conformal theory on $K3$ with an
extra $Z_{2}$ symmetry $g$.
It is precisely this $Z_{2}$ symmetry $g$ that we must orbifold
by, to freeze the $K3$ at the enhanced gauge symmetry point.

In the context of the main example we have been discussing, the
$C_i$ are the curves which shrink to zero size
as the heterotic string develops its $G^{H}\times E_{7}$ gauge symmetry.
We should orbifold the type IIA side by a combination
of orientifolding and simultaneous
action with $g$, to construct the dual to the heterotic theory.

What does $g$ correspond to on the world sheet?  One way
to study this limit is by making use of a linear sigma model
description of the model, following \phases\edcomments.
In this formalism, the Kahler modes corresponding to the
sizes of the $C_i$ are represented on the worldsheet by
$U(1)$ gauge multiplets, with the coefficients
$\vec r_i=(r_i^0,r_i^1,r_i^2)$ of (generalized)
worldsheet Fayet-Iliopoulos $D$-terms giving the sizes of
the $C_i$.  Each gauge multiplet $\Sigma_i$ contains four scalars
$\sigma_i$ which couple to charged hypermultiplets $\phi^i_\alpha$,
$\tilde\phi^i_\alpha$.
The bosonic potential is \edcomments
\eqn\bospotws{\eqalign{V=
&{1\over{2e^2}}
\sum_i\biggl\{\biggl(\bigl[\sum_\alpha Q_i^\alpha(|\phi^i_\alpha|^2
-|\tilde\phi^i_\alpha|^2)\bigr]-r_i^0\biggr)^2\cr
&+\biggl(Re(\sum_\alpha\phi^i_\alpha\tilde\phi^i_\alpha)-r_i^1\biggr)^2
+\biggl(Im(\sum_\alpha\phi^i_\alpha\tilde\phi^i_\alpha)-r_i^2\biggr)^2
\biggr\}\cr
&+{1\over 2}\sum_i\bigl[\sum_\alpha Q^{\alpha~2}_i
(|\phi^i_\alpha|^2+|\tilde\phi^i_\alpha|^2)\bigr]|\sigma_i|^2\cr}}

We see from \bospotws\ that for $r\rightarrow 0$ (with in addition
the worldsheet $\theta$ angles set to zero), the model develops
a singularity arising from the region of field space where
$\sigma_i\rightarrow\infty$ and $\phi^i_\alpha=0$.  Indeed,
this is the only regime where the model can be reliably studied
semiclassically for $r_i=\theta_i=0$.  In this limit,
the action $\Sigma\rightarrow -\Sigma$ on the worldsheet
gauge multiplet becomes a symmetry of the worldsheet model (combined
with $\phi\leftrightarrow -\tilde\phi$, though on the $\sigma$ branch all
the fields which couple to $\Sigma$ are hugely massive).
This appears to provide a worldsheet description of the geometrical action
$g$ described above.  Moreover, the vertex operators for the
Kahler blow-up modes are given by the fermions in the gauge multiplet
\silvwitt, so this action indeed projects out the Kahler deformations
corresponding to the $C_i$.
As we will
see below, the type II description of the heterotic dynamics
that we will present relies only on the details of the
action on the moduli and the action \acbase\ on the base of the $K3$
fibration.

\newsec{Black Hole Condensation and Supersymmetry Breaking}

Having seen in \S2
that the adiabatic argument allows us to construct N=1 dual pairs
with highly nontrivial dynamics expected on the heterotic side, we now
turn to a brief analysis of the dual descriptions of the infrared physics.

\subsec{Heterotic expectations}

Given the presence of a pure gauge group $G^{H}$, our expectation is that
gaugino condensation will occur.  In a globally supersymmetric theory, this
would lead to a mass gap and a discrete set of degenerate vacua \gaugecond.
The gaugino condensate is
\eqn\gc{\langle \chi_{\alpha}^{(b)}\chi^{(b)\alpha} \rangle \sim
\Lambda_{b}^{3} e^{i\gamma}}
where the phase $\gamma$ corresponds to a given discrete choice
of vacuum.  This will lead to a superpotential
\eqn\supgc{W = \sum_{b} h_{b} ~\Lambda_{b}^{3}(S)}
where
\eqn\group{G^{H} = \Pi ~G^{b}}
and where $b$ indexes the various hidden groups.  In \supgc\ we have noted that
in string theory the scale $\Lambda$ at which a given factor in $G^{H}$
becomes strong really depends on the dilaton $S$.
The constants $h_{b}$ include the phases $\gamma^{b}$ corresponding
to the discrete choices of vacua \drsw\krasnikov.

By analyzing the resulting bosonic potential (taking into account the
dependence of the Kahler potential $K$ and the superpotential $W$ on
all the chiral superfields in the low-energy theory at weak coupling), one
can in principle determine whether a stable vacuum exists at weak coupling
with broken supersymmetry and determine the resulting vacuum energy
(see \dixrev\kaplou\ for a review of various approaches to this problem).
Our problem is to understand how this structure is reproduced by the
type II dual, and to see whether the dual description
provides any useful insights about the physics, at least in particular
limits.

\subsec{Type II description}

Now we turn to the question of how the type II string can reproduce the
effects of gaugino condensation evident from the heterotic analysis.
There are several points we must take into
account, which will be crucial to the physics
on the type II side.  First of all,
the heterotic orbifold indicates the spectrum of the theory for
weak heterotic coupling, which is mapped to large
${\bf CP^1}$ radius $R$ on the type II side.
The purported nonperturbative vacuum
on the heterotic side is at small nonzero coupling,
corresponding to large finite $R$.
The orbifold acts freely on the ${\bf CP^1}$ base.
We are therefore
interested in the perturbative spectrum on the type II
side as a function of large finite $R$.
There are two features of the conifold locus at $R\rightarrow \infty$
which are crucial to understanding the type II physics:

\item{1)} The massless states in the $N=2$ theory
that are projected out by the $Z_2$
become invariant when given quantized internal momentum leading to
masses $\sim 1/R^2$ as
$R\rightarrow\infty$.
This implies that the full $N=2$ supersymmetry is restored as
$R\rightarrow\infty$.

\item{2)}The low-energy theory for type II at the conifold locus
contains massless solitonic states \strom\ in addition to
the perturbative states obtained from the orbifold.
One can see that these massless solitons survive the transition from the N=2
theory to the N=1 orientifold by examining the monodromies of the
gauge coupling functions \VW.

We will now take these points into account systematically.
The orbifold on the type II side
will have the same action
$(z_1,z_2)\rightarrow(\bar z_2,-\bar z_1)$ on the base
of the $K3$ fibration as it had on the base of the
elliptic fibration on the heterotic side.
This turns the ${\bf CP}^1$ into ${\bf RP}^2$,
which has nontrivial fundamental group
$\pi_1({\bf RP}^2)=Z_2$.
Therefore, a state that is projected out by the
orbifold will have a massive version, with appropriate momentum
along the nontrivial cycle $\gamma$, which is invariant under the $Z_2$.
More explicitly, in the adiabatic limit we can send the original
string state localized
along $\gamma$ with momentum $p={1\over R}$.  Then if
the original vertex operator $V$ transforms
as $V\rightarrow -V$ under the $Z_2$, the state
\eqn\newvert{V'=e^{i{x\over R}}f(x_\perp)V }
will be invariant. In \newvert\ $x$ is the
coordinate along the nontrivial cycle of the ${\bf RP}^2$,
and $f(x_\perp)$ localizes the string along $\gamma$.
The orbifold takes $x\rightarrow x + \pi R$ so that
the momentum factor gives a compensating factor of $-1$
under the $Z_2$.

This means that we will have massive versions of all the
fields $a_{b,D}^i$, with mass proportional to $1/R^2$.
In addition, we will have the
monopole hypermultiplets $M^b_i$ and $\tilde M^b_i$, which
survive the orbifold on the type II side \VW.
Because $N=2$ supersymmetry is restored as
$R\rightarrow\infty$, the massive $a_{b,D}^i$ fields will
couple to the surviving monopole fields as in the
$N=2$ theory.\foot{This should also follow from
the appropriate computation of the coupling of the
vertex operator to the D-brane monopole \joe.}
As discussed in the work of Seiberg and
Witten \SW\ (see also \sun\son),
in the global limit the mass terms involve the gauge invariant global
coordinates $u^b_\alpha$ (where $\alpha$ indexes the
Casimirs of the Lie algebra) on the moduli space,
which are functions of the $N=2$ scalar superpartners of the gauge
bosons, $a_{b,D}^i$, $i=1,\dots,{\rm rank}(G^H_b)$.

This reasoning tells us that
our massive fields appear in the superpotential
in terms proportional to the quadratic Casimirs $u^{b}_{2}$ of the
hidden groups $G^{H}_{b}$.
In the string theory context, the
coupling constant is the heterotic dilaton field $S$, which maps
by duality to the chiral superfield $y$ containing the ${\bf CP^1}$ radius
$R$.  So in our
case $u^b_\alpha=u^b_\alpha(a_{b,D},y)$.
Therefore as $R\rightarrow \infty$
there is a superpotential which looks like
\eqn\supII{W_{II}=\sum_b\biggl(m_bu^b_2(a^i_{b,D},y)+
\sum_{i=1}^{r(b)}M^b_ia^i_{b,D}\tilde M^b_i\biggr)}
on the type II side.  Here
$r(b)$ is the rank of
the $b$th factor in the hidden gauge group.

In analyzing the resulting
bosonic potential, we will (i) reproduce the general structure
of the potential arising from gaugino condensation on the
heterotic side and (ii) discover that the monopole fields
have vacuum expectation values, suggesting a geometric
description by analogy with the conifold transitions that
occur in the $N=2$ context \hubsh\GMS.

Before analyzing the physics of \supII, it is helpful to remember the
simplest case discussed in \SW.  In order to recover the physics of the
N=1 SU(2) theory from their solution of the N=2 SU(2) theory, Seiberg
and Witten perturb the N=2 theory by a superpotential which gives a mass to
the adjoint scalar in the N=2 vector multiplet.
Taken together with the couplings of the monopole fields $M, \tilde M$
which become massless at special points this implies
\eqn\swpert{W = mU(a_{D}) +  {\sqrt 2} a_{D} M\tilde M~.}
Then from the equations of motion together with the condition of
D-flatness they find that
\eqn\monvev{|\langle M \rangle| = |\langle \tilde M \rangle| = ({{-mU^{
\prime}(0)}\over {\sqrt {2}}})^{1/2} \neq 0~.}
The monopoles condense and give a mass to the (dual) U(1) gauge field by the
magnetic Higgs mechanism.  The resulting low energy theory has a gap -- this
is the dual explanation of confinement by monopole condensation.

Our expectation in the context of heterotic/type II duality is that the
superpotential \supII\ of our type II duals will give an analogous
picture, with light black hole condensation providing the dual
description of gaugino condensation
and supersymmetry breaking.
To make this more concrete, we must compute the bosonic potential
\eqn\potgen{V=e^K\biggl(D_iWG^{i\bar j}D_{\bar j}W-3|W|^2\biggr)+
{1\over 2}g^2D^2}
in terms of the Kahler potential $K$ and superpotential $W$.

Let us expand the superpotential \supII\ in $a^i_{b,D}$ in anticipation
of finding a minimum at small $a^i_{b,D}$.
\eqn\expandu{u_2(a^i_{b,D},y)=e^{i\gamma^{(b)}}\Lambda_b^2(y)+
{{\partial u_2}\over {\partial a^i_{b,D}}}a^i_{b,D}+\dots}
Recall also that the matching between the high and low-energy
theories gives the relation
\eqn\match{m_b\Lambda_{{\rm b,~high}}^2=\Lambda_{{\rm b,~low}}^3}
so we obtain a superpotential
\eqn\expandW{W_{II}=\sum_b\biggl(e^{i\gamma^{(b)}}\Lambda_b^3+
{{\partial u^b_2(y)}\over{\partial a_{b,D}^i}}a^i_{b,D}+
\sum_{i=1}^{r(b)}M^b_ia^i_{b,D}\tilde M^b_i\biggr)}
Since we are interested in comparing to the heterotic side,
let us work now in terms of the heterotic coupling $S$ and consider
$\Lambda_b$'s dependence on $S$.
Then we obtain the following bosonic potential
\eqn\bospot{\eqalign{V=
& e^K\biggl[\sum_{b,i(b)}
\bigl|h_bm_bu^b_{2,i}(S)+M^b_i\tilde M^b_i+K_iW\bigr|^2\cr
&+\bigl|\sum_b[h_b{{\partial\Lambda_b^3(S)}\over{\partial S}}
+h_b{{\partial u^b_{2,i}(S)}\over{\partial S}}a_{b,D}^i]+K_SW\bigr|^2\cr
&+\sum_b\sum_{i,\bar j=1}^{r(b)}
a_{b,D}^iM^b_iG^{\tilde M^b_i\bar{\tilde M^b_j}}
\bar a_{b,D}^j\bar M^b_j
+\sum_b\sum_{i,\bar j=1}^{r(b)}a_{b,D}^i\tilde M^b_iG^{M^b_i\bar{M^b_j}}
\bar a_{b,D}^j\bar{\tilde M^b_j}\cr
&+{\rm F-terms~of~other~fields}\cr
&-3\biggl|\sum_b\sum_{i=1}^{r(b)}
\bigl(h_b\Lambda_b^3(S)+h_bu^b_{2,i}(S)a_{b,D}^i
+M^b_ia_{b,D}^i\tilde M^b_i\bigr)+{\rm other~fields}\biggr|^2\biggr] \cr
&+{1\over{S_{II}+\bar S_{II}}}\sum_b\sum_{i(b)=1}^{r(b)}
\biggl(|M^b_i|^2-|\tilde M^b_i|^2\biggr)\cr}}
where we have used the notation
${{\partial u^b_2}\over{\partial a_{b,D}^i}}\equiv u^b_{2,i}$.

This expression of course depends on the Kahler potential.
In general, one would expect the Kahler potential to receive significant
loop (and nonperturbative) corrections, but duality allows us to work at
arbitrarily weak coupling on the type II side--given the
absence of field theoretic nonperturbative effects on the
type II side--unless
{\it stringy} nonperturbative effects
fix the type II dilaton away from weak coupling.
Assuming any such additional effects leave a minimum
at weak type II coupling, the tree level Kahler
potential should be a good approximation.\foot{If this assumption is false,
the duality we discuss still applies but the analysis of
the bosonic potential changes accordingly to take into
account its dependence on the moduli coming from
$N=2$ hypermultiplets.}
The vacuum will also depend
on the contributions to the bosonic potential
of fields other than those on which
we are focusing.  Different sets of assumptions
and methods for analyzing the potential
exist in the literature (see for example \dixrev\kaplou\
and references therein).

Assuming there is a minimum near the $a_{b,D}^i=0$ minimum
of the rigid case \swpert, the monopole fields $M$ and $\tilde M$ will
minimize the first term in \bospot\ (consistent with D-flatness)
up to supergravity corrections.
Setting $D_{i}W = 0$ yields
\eqn\montwo{\langle M^b_i\tilde M^b_i \rangle = -h_bm_bu^b_{2,i}(S)-K_iW}
Here the first term is in accord with \monvev\
and the second term arises from supergravity.
Now having
integrated out the extra particles $M$, $\tilde M$, and
$a_D$, we obtain the same general form of bosonic potential
as arises from the heterotic side.  We
must then minimize with respect to the dilaton and
all the other scalars in the model.
Then supersymmetry is broken if there is any field $\phi$
for which
\eqn\Fterm{F_\phi=e^{{K\over 2}}G^{\phi\bar\phi}
\biggl({{\partial W}\over {\partial \phi}}+
{{\partial K}\over {\partial \phi}}W\biggr)\ne 0}
in the vacuum.

As in the global
situation analyzed in \SW, the dual description of the
effects of gaugino condensation involves a mass perturbation
breaking the $N=2$ supersymmetry.  We have seen that
the orbifold produces the necessary massive mode as a Kaluza Klein
excitation of the original variable $u_2^{b}$ that was projected
out.

One intriguing feature of the type IIA vacuum is the presence of
nonzero vacuum expectation values of the
monopole fields (wrapped two-branes) $M$ and $\tilde M$.
In the context of $N=2$ compactifications
of the type II string theory, such vacuum expectation values can
be turned on continuously when consistent with
D- and F-flatness, giving transitions to other branches
of the moduli space \GMS.  There is a well-known geometrical description
of the conformal field theories involved in this process \hubsh.
For example in type IIB string theory one
approaches the conifold locus in complex structure
moduli space by deformations causing
appropriate $S^3$s to shrink to zero size.
One can then resolve this singularity by replacing the tips
of the resulting cones by ${\bf CP^1}\sim S^2$s.

At the generic conifold such a ``small resolution'' does
not produce a Kahler manifold \nonkahlerres.  This
was noted by Candelas, de la Ossa, Green, and Parkes, who speculated that
such resolutions may correspond to
supersymmetry-breaking directions \cdgp.
The analysis presented here suggests a realization of
these ideas through duality.
By analogy with the quantum conifold transitions in the
$N=2$ context, we expect that the nonzero monopole
VEVs we have found correspond to a vacuum which has a conformal field theory
description involving strings propagating on the non-Kahler
resolutions of conifold singularities.  If this analogy
holds, then the fact that the corresponding conformal
field theory is nonsupersymmetric might in fact provide the simplest
method for establishing that supersymmetry is broken in such
theories.

Given a conformal field theory description,
a very general argument suggests
that the leading approximation (in the type II coupling) to the
cosmological constant $\it must$ vanish in this class
of theories.
Although on the heterotic side detailed dynamical assumptions are usually
invoked, on the type II side this statement follows simply from
the fact that the leading
contributions to the vacuum energy
vanish by $SL(2,C)$
invariance.

\newsec{Conclusions}

We have seen that the construction of dynamical
4d N=1 dual pairs is possible
by application of the adiabatic argument of \VW.
In particular, one can realize ``racetrack models'' of supersymmetry
breaking in a dual type II description.

Our analysis suggests that
the tree-level superpotential on the type
II side reproduces the bosonic potential expected from
gaugino condensation on the heterotic side, by generalizing the mechanism
of \SW\ which explained the gap of pure N=1 SU(2) gauge theory.
A careful study of the geometry of the
type II compactification (including possibly the non-Kahler
resolutions of the conifold singularity) might therefore translate
into detailed information about the mechanism of supersymmetry
breaking in these models.  The $N=2$ models we have started with here
are rather cumbersome, as they contain numerous moduli.  It
would be nice to find simpler examples of this
phenomenon which can be more easily studied in detail.
More generally, one would like to extend the class of
useful $N=1/N=0$ dualities
to models which are not obviously obtained as orbifolds
of $N=2$ dual pairs.

\centerline{\bf Acknowledgements}
\nobreak
We are grateful to P. Aspinwall,
C. Vafa, E. Witten, and especially D. Morrison
for very helpful discussions.  S.K.
is supported by
the Harvard Society of Fellows and the William F. Milton
Fund of Harvard University. The research of E.S. is
partially supported by a grant from the AT$\&$T GRPW.

\listrefs
\end